\documentclass[showpacs,showkeys,twoside,eqsecnum,twocolumn]{revtex4} 

\usepackage{amsfonts,amsmath,amssymb,bm,hyperref,empheq,graphicx,dsfont}

\begin{document}


\title{Neutrino oscillations in matter and in twisting magnetic fields}

\author{Maxim Dvornikov}
\affiliation{Department of Physics, P.O. Box 35, FIN-40014, University of
Jyv\"{a}skyl\"{a}, Finland}
%
\affiliation{IZMIRAN, 142190, Troitsk, Moscow region, Russia} 

\date{\today}

\begin{abstract}
We find the solution to the Dirac equation for a massive neutrino with a magnetic 
moment propagating in background matter and interacting with the twisting magnetic 
field. In frames of the relativistic quantum mechanics approach to the description 
of neutrino evolution we use the obtained solution to derive neutrino wave 
functions satisfying the given initial condition. We apply the results to the 
analysis of neutrino spin oscillations in matter under the influence of the 
twisting magnetic field. Then on the basis of the yielded results we describe 
spin-flavor oscillations of Dirac neutrinos that mix and have non-vanishing matrix 
of magnetic moments. We again formulate the initial condition problem, derive 
neutrinos wave functions and calculate the transition probabilities for different 
magnetic moments matrices. The consistency of the obtained results with the quantum 
mechanical treatment of spin-flavor oscillations is discussed. We also consider 
several applications to astrophysical and cosmological neutrinos.  
\end{abstract}



\maketitle

\section{Introduction}

Neutrino conversions from one flavor to another combined with the change of the 
particle helicity, e.g. $\nu_e^\mathrm{L} \leftrightarrow \nu_\mu^\mathrm{R}$, are 
usually called neutrino spin-flavor oscillations (see Ref.~\cite{qmSFO}). This 
neutrino oscillations scenario is important since it could be the explanation of 
the time variability of the solar neutrino flux (see, e.g., Ref.~\cite{solarnu}). 
Massive flavor neutrinos are known to mix and can have non-zero magnetic moments. 
The influence of the strong magnetic field with the realistic profile could lead to 
the spin-flavor oscillations of solar neutrinos (see, e.g., 
Ref.~\cite{realisticB}). Moreover, studying neutrino spin-flavor oscillations 
happening inside the Sun, one will be able to discriminate between different solar 
models~\cite{PicPulAndBarMan07}. However it was found out in Ref.~\cite{smallcontr} 
that neutrino spin-flavor oscillations in solar magnetic fields give a sub-dominant 
contribution to the total conversion of solar neutrinos.

In this paper we study neutrino spin and spin-flavor oscillations in matter and in 
an external magnetic field. We suppose that a neutrino is a Dirac particle with a 
non-zero magnetic moment. It should be mentioned that in spite of the recent claims 
of the experimental confirmation that neutrinos are Majorana 
particles~\cite{KleKriDieChk04}, the question about the nature of neutrinos is 
still open~\cite{EllEng04}. The possibility to distinguish between Dirac and 
Majorana particles in the partially polarized solar neutrino flux, due to the 
spin-flavor precession, was examined in Ref.~\cite{Sem97}. 

To describe the evolution of the neutrino system we apply the technique based on 
the relativistic quantum mechanics. We start from the exact solution to the Dirac 
equation in an external field and then derive the neutrino wave functions 
satisfying the given initial condition. We used this method to describe neutrino 
flavor oscillations in vacuum~\cite{FOvac}, in background matter~\cite{Dvo06EPJC} 
and spin-flavor oscillations in an external magnetic field~\cite{DvoMaa07}. Note 
that neutrino spin-flavor oscillations in electromagnetic fields of various 
configurations were examined in Refs.~\cite{emfields,Dvo07YadFiz,twisting} using 
the standard quantum mechanical approach.


In Sec.~\ref{SO} we find the solution to the Dirac equation for a neutrino 
propagating in background matter and interacting with the twisting magnetic field. 
Then we formulate the initial condition problem and receive the transition 
probability for spin oscillations in the given external fields. The standard 
quantum mechanical transition probability formula is re-derived and the conditions 
of its validity are analyzed. In Sec.~\ref{SFO} we apply the obtained Dirac 
equation solutions to the description of neutrino spin-flavor oscillations in the 
twisting magnetic field. First we discuss magnetic moment matrices of neutrinos in 
flavor and mass eigenstates bases. Then we solve the initial condition problem in 
two different cases of the magnetic moments matrix in the mass eigenstates basis 
with (i) great diagonal elements and (ii) great non-diagonal elements. Note that 
the analogous magnetic moments matrices were discussed in Ref.~\cite{DvoMaa07}. We 
get neutrinos wave functions and calculate transition probabilities for processes 
like $\nu_\beta^\mathrm{L}\xrightarrow{B}\nu_\alpha^\mathrm{R}$. The consistency of 
the Dirac-Pauli equation approach with the standard quantum mechanical treatment of 
spin-flavor oscillations, based on the Schr\"odinger evolution equation, is 
considered in Sec.~\ref{QM}. Then in Sec.~\ref{APPL} we present some applications 
and finally we summarize our results in Sec.~\ref{CONCL}.


\section{Neutrino spin oscillations in matter and in a twisting magnetic 
field\label{SO}}

In this section we obtain the exact solution to the Dirac-Pauli equation for a 
neutrino interacting with background matter and a twisting magnetic field and 
discuss spin oscillations of a single Dirac neutrino in the given external fields. 

A neutrino is taken to have the non-zero mass $m$ and the magnetic moment $\mu$. 
The Lagrangian for this system has the form,
\begin{equation}\label{LagrmattB}
  \mathcal{L}=\bar{\nu}(\mathrm{i}\gamma^\mu\partial_\mu - m)\nu-
  \bar{\nu}\gamma_\mu^\mathrm{L}\nu f^\mu-
  \frac{\mu}{2}\bar{\nu}\sigma_{\mu\nu}\nu F^{\mu\nu},
\end{equation}
where $\gamma_\mu^\mathrm{L}=\gamma_\mu(1+\gamma^5)/2$,
$\sigma_{\mu\nu}=(\mathrm{i}/2)(\gamma_\mu\gamma_\nu-\gamma_\nu\gamma_\mu)$
and $F_{\mu\nu}=(\mathbf{E},\mathbf{B})$ is the electromagnetic
field tensor. In the following we will discuss the situation when
only magnetic field $\mathbf{B}$ is presented, i.e.
$\mathbf{E}=0$. The neutrino interaction with matter is
characterized by the four vector $f^\mu$. For the non-moving and
unpolarized matter one can take that the spatial components of the
vector $f^\mu$ are zero, i.e. $\mathbf{f}=0$. If, for instance, we consider an
electron neutrino propagating in matter, which consists of
electrons, protons and neutrons, we obtain for the time component, $f^0$,
of the vector $f^\mu$ (see, e.g., Ref.~\cite{DvoStu02JHEP}),
\begin{align}\label{fcomp}
  f^0 = & \sqrt{2}G_\mathrm{F}
  \sum_{f=e,p,n} n_f q_f,
  \notag
  \\
  q_f = & (I_{3\mathrm{L}}^{(f)}-2Q^{(f)}\sin^2\theta_W+\delta_{ef}),
\end{align}
where $n_f$ is the number density of background particles, $I_{3\mathrm{L}}^{(f)}$ 
is the third isospin component of the matter
fermion $f$, $Q^{(f)}$ is its electric charge, $\theta_{W}$ is
the Weinberg angle and $G_\mathrm{F}$ is the Fermi constant.

It should be noted that Eqs.~\eqref{LagrmattB} and~\eqref{fcomp} constitute the 
phenomenological model studied in the present paper. These expressions are valid in 
a relatively weak external magnetic field. For example, one has to take into 
account the spatial components of the vector $f^\mu$ if we describe neutrino 
propagation in background matter composed of electrons under the influence of a 
very strong magnetic field with $\sqrt{|\mathbf{B}|} \gg 
\max(m_e,T,\mathfrak{M},|\mathbf{p}|)$, where $m_e$ is the electron mass, $T$ is 
the temperature of background matter, $\mathfrak{M}$ is its chemical potential and 
$\mathbf{p}$ is the neutrino momentum. This situation was analyzed in 
Ref.~\cite{EliFerInc04}.    

Using Eq.~\eqref{LagrmattB} one writes down the Dirac equation which accounts for 
the neutrino interaction with matter and magnetic field,
\begin{align}\label{DireqmattB}
  \mathrm{i}\dot{\nu}= & \mathcal{H}\nu,
  \notag
  \\
  \mathcal{H}= & (\bm{\alpha}\hat{\mathbf{p}})+\beta m -
  \mu\beta(\bm{\Sigma}\mathbf{B})+f^0(1+\gamma^5)/2,
\end{align}
where $\bm{\alpha}=\gamma^0\bm{\gamma}$, $\beta=\gamma^0$ and 
$\bm{\Sigma}=\gamma^0\gamma^5\bm{\gamma}$ are the Dirac matrices. Let us discuss 
the case of the twisting magnetic field, $\mathbf{B}=B(0,\sin \omega x,\cos \omega 
x)$, where $\omega$ is the frequency of the magnetic field rotation. Sometimes it 
is called the spiral undulator magnetic field. Note that neutrino oscillations in 
twisting magnetic fields in frames of the quantum mechanical approach were studied 
in Ref.~\cite{twisting}.

We notice that the Hamiltonian $\mathcal{H}$ in
Eq.~\eqref{DireqmattB} depends on neither $y$ nor $z$ coordinates.
Therefore we assume that the wave function depends on these
coordinates exponentially, $\nu \sim \exp(\mathrm{i}p_y y +
\mathrm{i}p_z z)$, where $p_y$ and $p_z$ are constant values. Then
for simplicity one can take that $p_y=p_z=0$. It means that a
neutrino moves along the undulator axis. Let us express the neutrino wave function 
in terms of the two
component spinors, $\nu^\mathrm{T}=(\varphi,\chi)$. On the basis
of Eq.~\eqref{DireqmattB} we receive equations for the two
component spinors,
\begin{align}\label{phichi}
  \mathrm{i}\dot{\varphi} = &
  (m-\mu(\bm{\sigma}_{\perp{}}\mathbf{B})+f^0/2)\varphi +
  (\sigma_1 \hat{p}_x-f^0/2) \chi,
  \notag
  \\
  \mathrm{i}\dot{\chi} = &
  (-m+\mu(\bm{\sigma}_{\perp{}}\mathbf{B})+f^0/2)\chi +
  (\sigma_1 \hat{p}_x-f^0/2) \varphi,
\end{align}
where $\hat{p}_x=-\mathrm{i}\partial_x$ and
$\bm{\sigma}_{\perp{}}=(\sigma_2,\sigma_3)$.

Now we replace the neutrino wave function $\nu$ with the new one, $\nu \to 
\tilde{\nu} = \mathcal{U}^\dag \nu$, where 
$\mathcal{U}=\mathrm{diag}(\mathfrak{U},\mathfrak{U})$ and 
$\mathfrak{U}=\cos(\omega x/2)+\mathrm{i}\sigma_1\sin(\omega x/2)$. Then we again 
express the new wave function using the two component spinors, 
$\tilde{\nu}^\mathrm{T}=(\xi,\eta)$, with $\varphi=\mathfrak{U}\xi$ and 
$\chi=\mathfrak{U}\eta$. With help of the following properties of the matrix 
$\mathfrak{U}$:
$\mathfrak{U}^\dag(\bm{\sigma}_{\perp{}}\mathbf{B})\mathfrak{U}=\sigma_3 B$, 
$\mathrm{d}\mathfrak{U}/\mathrm{d}x=\mathrm{i}\sigma_1 \omega \mathfrak{U}/2$ and 
$\mathfrak{U}^\dag \sigma_1 \mathfrak{U}=\sigma_1$, as well as using 
Eq.~\eqref{phichi} we arrive to the equations for the new two component spinors,
\begin{align}\label{xieta}
  \mathrm{i}\dot{\xi} = &
  (m-\mu B \sigma_3+f^0/2)\xi 
  \notag
  \\
  & + [(\omega-f^0)/2+\sigma_1 \hat{p}_x]\eta,
  \notag
  \\
  \mathrm{i}\dot{\eta} = &
  (-m+\mu B \sigma_3+f^0/2)\eta 
  \notag
  \\
  & + [(\omega-f^0)/2+\sigma_1 \hat{p}_x]\xi.
\end{align}
We notice that Eq.~\eqref{xieta} do not contain the dependence on $x$ coordinate. 
Thus one gets that the new wave function depends on
$x$ as $\tilde{\nu} \sim \exp(\mathrm{i}p x)$, where $p$ is a
constant value, the analog of the particle momentum. It means that we can
replace $\hat{p}_x \to p$ in Eq.~\eqref{xieta}.

We look for stationary solutions to Eq.~\eqref{xieta}, i.e. $\tilde{\nu} \sim 
\exp(-\mathrm{i}E t)$. Supposing that this equation has a non-trivial solution we 
receive the energy levels in the form, $E = f^0/2 \pm E^{(\zeta)}$. The function 
$E^{(\zeta)}$ depends on
the momentum and the characteristics on the external fields as
\begin{equation}\label{EnergymattB}
  E^{(\zeta)}=\sqrt{\mathcal{M}^2+m^2+p^2-2\zeta R^2},
\end{equation}
where $R^2=\sqrt{p^2 \mathcal{M}^2 + (\mu B)^2 m^2}$ and $\mathcal{M}=\sqrt{(\mu 
B)^2 + (\omega-f^0)^2/4}$.
In Eq.~\eqref{EnergymattB} $\zeta=\pm 1$ is the discrete quantum number.

Using energy spectrum~\eqref{EnergymattB} we can reproduce the results of the 
previous works where the Dirac equation for a neutrino interacting with various 
external fields was solved. Namely,
\begin{itemize}
  \item neutrino interaction with a constant transversal magnetic field
  (see, e.g., Ref.~\cite{DvoMaa07}). This situation corresponds to
  $\omega=0$ and $f^0=0$.
  Using Eq.~\eqref{EnergymattB} we get
  $E = \pm \left( \sqrt{m^2+p^2}-\zeta\mu B \right)$ that coincides with the
  energy spectrum used in Ref.~\cite{DvoMaa07};
  \item neutrino interaction with background matter
  (see Ref.~\cite{matterQFT}). This case corresponds to $\omega=0$ and
  $B=0$. With help of Eq.~\eqref{EnergymattB} we receive that
  $E = f^0/2 \pm \sqrt{(p-\zeta f^0/2)^2 + m^2}$ that coincides with
  the results of Ref.~\cite{matterQFT}.
\end{itemize}
Note that, if we set $\omega=0$ and $B \neq 0$ in Eq.~\eqref{EnergymattB}, we 
arrive to the case of a neutrino propagating in background matter under the 
influence of a constant transversal magnetic field.

The basis spinors $u^{(\zeta)}$ and $v^{(\zeta)}$ corresponding to the signs 
$\pm{}$ in the dispersion relation can be found from Eq.~\eqref{xieta}. The general 
expressions for these spinors, which account for the particle mass exactly, are 
rather complicated. Therefore we present here the basis spinors for a relativistic 
neutrino with $(m/E) \ll 1$,
\begin{align}\label{spinorsmattB}
  u^{(\zeta)}= &
  \frac{1}{2\sqrt{2\mathcal{M}(\mathcal{M}+\varDelta)}}
  \begin{pmatrix}
     \mu B+\zeta\mathcal{M}+\varDelta \\
     \mu B-\zeta\mathcal{M}-\varDelta \\
     \mu B-\zeta\mathcal{M}-\varDelta \\
     \mu B+\zeta\mathcal{M}+\varDelta \
  \end{pmatrix},
  \notag
  \\
  v^{(\zeta)}= &
  \frac{1}{2\sqrt{2\mathcal{M}(\mathcal{M}-\varDelta)}}
  \begin{pmatrix}
     \mathcal{M}-\varDelta-\zeta\mu B \\
     \mathcal{M}-\varDelta+\zeta\mu B \\
     \varDelta-\mathcal{M}-\zeta\mu B \\
     \varDelta-\mathcal{M}+\zeta\mu B \
  \end{pmatrix},
\end{align}
where $\varDelta = (\omega-f^0)/2$. Note that the basis spinors in 
Eq.~\eqref{spinorsmattB} satisfy the orthonormality conditions,
\begin{equation}\label{oncso}
  u^{(\zeta)\dag}u^{(\zeta')}=v^{(\zeta)\dag}v^{(\zeta')}=
  \delta_{\zeta\zeta'},
  \quad
  u^{(\zeta)\dag}v^{(\zeta')}=0.
\end{equation}

\subsection{Neutrino evolution in matter under the influence of a twisting magnetic 
field}

Using the approach developed in our previous works~\cite{FOvac,Dvo06EPJC,DvoMaa07} 
we can formulate the initial condition problem for the system in question. For the 
given initial wave function $\nu(x,0)$ one should find the wave function $\nu(x,t)$ 
at subsequent moments of time, while a particle propagates in the external
fields. This wave function has the form (see
Refs.~\cite{FOvac,Dvo06EPJC,DvoMaa07}),
\begin{equation}\label{numattB}
  \nu(x,t)=
  \mathcal{U}(x)
  e^{-\mathrm{i} f^0 t/2}
  \int_{-\infty}^{+\infty}\frac{\mathrm{d}p}{2\pi}
  e^{\mathrm{i}px}S(p,t)\tilde{\nu}(p,0),
\end{equation}
where
\begin{equation}\label{FTtildenu}
  \tilde{\nu}(p,0)=
  \int_{-\infty}^{+\infty}\mathrm{d}x
  e^{-\mathrm{i}px}\mathcal{U}^\dag(x)\nu(x,0),
\end{equation}
is the Fourier transform of the initial condition for the fermion $\tilde{\nu}$ and
\begin{align}\label{PJmattB}
  S(p,t)= &
  \sum_{\zeta=\pm 1}
  \bigg[
    \left(
      u^{(\zeta)}\otimes u^{(\zeta)\dag}
    \right)
    \exp{(-\mathrm{i}E^{(\zeta)} t)}
	\notag
	\\
	& +
    \left(
      v^{(\zeta)}\otimes v^{(\zeta)\dag}
    \right)
    \exp{(+\mathrm{i}E^{(\zeta)} t)}
  \bigg],
\end{align}
is the analog for the Pauli-Jourdan function for a spinor field interacting with 
matter and a twisting magnetic field. The basis spinors $u^{(\zeta)}$ and 
$v^{(\zeta)}$ are presented in Eq.~\eqref{spinorsmattB}. To derive 
Eqs.~\eqref{numattB}-\eqref{PJmattB} we use orthonormality of the basis 
spinors~\eqref{oncso}.

Let us suppose that initially a neutrino is in the state with the following wave 
function: $\nu(x,0)=e^{\mathrm{i}kx}\xi_0$, where 
$\xi_0^\mathrm{T}=(1/2)(1,-1,-1,1)$. It is possible to check that 
$(1/2)(1-\Sigma_1)\xi_0=\xi_0$. Hence, the spinor $\nu(x,0)$ describes a particle 
propagating along the $x$-axis, with its spin directed opposite to the $x$-axis, 
i.e. a left-handed neutrino. Analogous initial condition was adopted in 
Refs.~\cite{FOvac,Dvo06EPJC,DvoMaa07} where neutrino flavor and spin-flavor 
oscillations were studied. 

Using Eq.~\eqref{FTtildenu} we find that $\tilde{\nu}(p,0)=2\pi 
\delta(p-k-\omega/2) \xi_0$. It is interesting to note that the following identity 
is satisfied: $\left( v^{(\zeta)}\otimes v^{(\zeta)\dag} \right)\xi_0 = 0$. 
Therefore no particles with "negative" energies appear in neutrino interacting with 
considered external fields. Using Eqs.~\eqref{spinorsmattB} 
and~\eqref{numattB}-\eqref{PJmattB} as well as the chosen initial condition we 
arrive to the right-polarized component of the final wave function,
\begin{align}\label{nuR}
  \nu^\mathrm{R}(x,t)= &
  \frac{1}{2}(1+\Sigma_1)\nu(x,t)
  \\
  = &
  \exp[\mathrm{i}(k+\omega)x-\mathrm{i} f^0 t/2]
  \notag
  \\
  & \times
  \notag
  \frac{\mu B}{2\mathcal{M}}
  \left.
  \left(
    e^{-\mathrm{i}E^{+{}}t}-e^{-\mathrm{i}E^{-{}}t}
  \right)
  \right|_{p=k+\omega/2}\kappa_0,
\end{align}
where $\kappa_0^\mathrm{T}=(1/2)(1,1,1,1)$.

Supposing that initially no right-polarized particles are present
and with help of Eq.~\eqref{nuR} we calculate the transition
probability for the process $\nu^\mathrm{L} \to \nu^\mathrm{R}$,
\begin{align}\label{PtrLR}
  P_{\nu^\mathrm{L} \to \nu^\mathrm{R}}(t)= &
  \frac{(\mu B)^2}{(\mu B)^2 + \varDelta^2}
  \notag
  \\
  & \times
  \sin^2
  \left.
  \left(
    \frac{E^{+{}}-E^{-{}}}{2}t
  \right)
  \right|_{p=k+\omega/2}.
\end{align}
It can be seen from Eq.~\eqref{PtrLR} that the resonance in neutrino spin 
oscillations occurs when $\varDelta \to 0$. One finds from Eq.~\eqref{EnergymattB} 
that $(E^{+{}}-E^{-{}})/2 = -\mu B$ at $\varDelta=0$. Therefore the resonance 
transition probability is always $P_\mathrm{res}(t)=\sin^2(\mu B t)$. 

To analyze Eq.~\eqref{PtrLR} we introduce the group velocity,
\begin{equation}\label{grvel}
  \mathcal{V}^{(\zeta)}=\frac{\partial E}{\partial p}=
  \frac{p}{E^{(\zeta)}}
  \left(
    1-\zeta\frac{\mathcal{M}^2}{R^2}
  \right).
\end{equation}
Now we can distinguish three different cases.
\begin{enumerate}
  \item \label{case1} 
  First we suppose that $p=0$. This situation can happen if 
  $k=-\omega/2$.  With
  help of Eq.~\eqref{EnergymattB} we obtain that the energy levels are 
  $E^{(\zeta)} = \sqrt{(m - \zeta \mu B)^2 + \varDelta^2}$. 
  Using Eq.~\eqref{grvel}
  we receive that the group velocity vanishes, $\mathcal{V}^{(\zeta)}=0$. It
  means that the neutrino is captured by the twisting magnetic field.
  \item \label{case2} 
  Now we assume that $p=\pm\mathcal{M}$, with $p\neq 0$. 
  For the definiteness we discuss the situation when $p=\mathcal{M}$ 
  since the case $p=-\mathcal{M}$ can be considered analogously.
  The energies corresponding to different values of $\zeta$ are
  \begin{align}\label{Epm2case}
    E^{+{}} = & m
    \sqrt{1-\frac{(\mu B)^2}{\mathcal{M}^2}
    +\frac{(\mu B)^4 m^2}{2\mathcal{M}^6}},
	\notag
	\\
    E^{-{}} = & 2\mathcal{M}
    \left(
      1+\frac{m^2}{8\mathcal{M}^2}
      \left[
        1+\frac{(\mu B)^2}{\mathcal{M}^2}
      \right]
    \right).
  \end{align}
  Using Eq.~\eqref{grvel} one can compute the group velocities,
  \begin{align}\label{Vpm2case}
    \mathcal{V}^{+{}} = &
    \frac{(\mu B)^2 m}{2\mathcal{M}^3}
    \left[
      1-\frac{(\mu B)^2}{\mathcal{M}^2}
      +\frac{(\mu B)^4 m^2}{2\mathcal{M}^6}
	\right]^{-1/2},
	\notag
	\\
	\mathcal{V}^{-{}} = & 1-
    \frac{m^2}{8\mathcal{M}^2}
    \left[
      1+3 \frac{(\mu B)^2}{\mathcal{M}^2}
    \right].
  \end{align}
  In Eqs.~\eqref{Epm2case} and~\eqref{Vpm2case} we suppose that 
  $m \ll \mathcal{M}$. On the basis of 
  Eqs.~\eqref{Epm2case} and~\eqref{Vpm2case} we get the resonance energies
  ($\varDelta \to 0$),
  \begin{align}
    E^{+{}}_\mathrm{res} \to & \frac{m^2}{\sqrt{2} \mu B},
    \notag
	\\
    E^{-{}}_\mathrm{res} \to & 2 \mu B
    \left[
      1+\frac{m^2}{4(\mu B)^2}
    \right],
  \end{align}
  and group velocities
  \begin{equation}\label{Vpm2caseRes}
    \mathcal{V}^{+{}}_\mathrm{res} \to \frac{1}{\sqrt{2}},
    \quad
    \mathcal{V}^{-{}}_\mathrm{res} \to 1-
    \frac{m^2}{2(\mu B)^2}.
  \end{equation}
  It should be noted that group velocities are always less than one,
  $\mathcal{V}^{\pm{}}<1$ [see, e.g., Eq.~\eqref{Vpm2caseRes}].
  %
  \item The last situation is realized when $p \neq \pm\mathcal{M}$ 
  and $p \neq 0$. The energies in this case have the form,
  \begin{align}\label{energy4case}
    E^{(\zeta)} = &
    p-\zeta\mathcal{M}
	\notag
	\\
	& +
    \frac{m^2}{2(p-\zeta\mathcal{M})}
    \left[
      1-\zeta\frac{(\mu B)^2}{\mathcal{M}p}
    \right].
  \end{align}
  The expression for the transition probability~\eqref{PtrLR} is now rewritten in 
  the following way:
  \begin{align}\label{PtrLR4case}
    P(t)= & \frac{(\mu B)^2}{(\mu B)^2 + \varDelta^2}
	\notag
	\\
	& \times
	\sin^2
    \left(
      \sqrt{(\mu B)^2 + \varDelta^2}t
    \right).
  \end{align}
  Note that transition probability expressions for spin oscillations derived 
  earlier (see, e.g., Ref.~\cite{twisting}) coincide with 
  Eq.~\eqref{PtrLR4case} which is valid only if $p \neq \pm\mathcal{M}$ 
  and $p \neq 0$.
\end{enumerate}

It should be noted that the "non-standard" regimes in neutrino spin oscillations 
described in items~\ref{case1} and~\ref{case2} are likely to be realized for 
neutrinos with small initial momenta (see also Sec.~\ref{APPL} below). 

\section{Neutrino spin-flavor oscillations in a twisting magnetic field\label{SFO}}

Now we apply the results of the previous section to the description of neutrino 
spin-flavor oscillations in a twisting magnetic field. Let us study the evolution 
of two Dirac neutrinos $(\nu_\alpha,\nu_\beta)$ that mix and interact with the 
external electromagnetic field $F_{\mu\nu}$. The Lagrangian for this system has the 
form
\begin{align}\label{Lagrnu}
  \mathcal{L}(\nu_{\alpha},\nu_{\beta})= &
  \sum_{\lambda=\alpha,\beta}\bar{\nu}_\lambda 
  \mathrm{i}\gamma^\mu\partial_\mu \nu_\lambda -  
  \sum_{\lambda\lambda'=\alpha,\beta}
  \bigg[
    m_{\lambda\lambda'} \bar{\nu}_\lambda \nu_{\lambda'}
	\notag
    \\
    & +
	\frac{1}{2}	 
	M_{\lambda\lambda'}
	\bar{\nu}_{\lambda}\sigma_{\mu\nu}\nu_{\lambda'} F^{\mu\nu}
  \bigg],
\end{align}
Here $(m_{\lambda\lambda'})$ and $(M_{\lambda\lambda'})$ are the mass and the 
magnetic moments matrices that are generally independent. By definition these 
matrices are intoduced in the flavor eigenstates basis. The electromagnetic field 
is taken to have the same configuration as in Sec.~\ref{SO}.

To analyze the dynamics of the system we again set the initial condition by 
specifying the initial wave functions of the flavor neutrinos $\nu_{\lambda}$ and 
then analytically determine the field distributions at following moments of time. 
We assume that the initial condition is
\begin{equation}\label{inicondnu}
  \nu_{\alpha}(x,0)=0,
  \quad
  \nu_{\beta}(x,0)=\xi(x),
\end{equation}
where $\xi(x)$ is a function to be specified. One of the possible choices for the 
initial condition for $\nu_\beta$ is the plane wave field distribution, 
$\xi(x)=e^{\mathrm{i} k x}\xi_0$ (see Refs.~\cite{FOvac,Dvo06EPJC,DvoMaa07}). If we 
study ultrarelativistic initial particles, we can choose the spinor $\xi_0$ as in 
Sec.~\ref{SO}, i.e. in the following form: $\xi_0^\mathrm{T}=(1/2)(1, -1, -1, 1)$.

In order to eliminate the vacuum mixing term in Eq.~\eqref{Lagrnu}, i.e. to 
diagonalize the mass matrix, we introduce a new basis of the wave functions, the 
mass eigenstate basis $\psi_a$, $a=1,2$, obtained from the original flavor basis 
$\nu_{\lambda}$ through the unitary transformation
\begin{equation}\label{matrtrans}
  \nu_{\lambda}=\sum_{a=1,2}U_{\lambda a}\psi_a,
\end{equation}
where the matrix $({U}_{\lambda a})$ is parametrized in terms of a mixing angle 
$\theta$ as usual
\begin{equation}\label{matrU}
  ({U}_{\lambda a})=
  \begin{pmatrix}
    \cos \theta & -\sin \theta \\
    \sin \theta & \cos \theta \
  \end{pmatrix}.
\end{equation}
The Lagrangian~\eqref{Lagrnu} rewritten in terms of the fields $\psi_a$ takes the 
form
\begin{align}\label{Lagrpsi}
  \mathcal{L}(\psi_1,\psi_2)= & \sum_{a=1,2}\mathcal{L}_0(\psi_a)
  \notag
  \\
  & -
  \frac{1}{2}
  \sum_{ab=1,2}\mu_{ab}\bar{\psi}_a\sigma_{\mu\nu}\psi_b F^{\mu\nu},
\end{align}
where $\mathcal{L}_0(\psi_a)=\bar{\psi}_a(\mathrm{i}\gamma^\mu 
\partial_\mu-m_a)\psi_a$ is the Lagrangian for the free fermion $\psi_a$ with the 
mass $m_a$ and
\begin{equation}\label{magmomme}
  \mu_{ab}=\sum_{\lambda\lambda'=\alpha,\beta}
  U^{-1}_{a\lambda}{M}_{\lambda\lambda'}U_{\lambda' b},
\end{equation}
is the magnetic moment matrix presented in the mass eigenstates basis. Using 
Eqs.~\eqref{inicondnu}-\eqref{matrU} the initial conditions for the fermions 
$\psi_a$ become
\begin{equation}\label{inicondpsi}
  \psi_1(x,0)=\sin\theta\xi(x),
  \quad
  \psi_2(x,0)=\cos\theta\xi(x).
\end{equation}

For the given configuration of the electric and magnetic fields we write down the 
Dirac-Pauli equation for $\psi_a$, resulting from Eq.~\eqref{Lagrpsi}, as follows:
\begin{equation}\label{Direqpsi}
  \mathrm{i}\dot{\psi}_a=\mathcal{H}_a\psi_a+V\psi_b,
  \quad
  a,b=1,2,
  \quad
  a \neq b,
\end{equation}
where $\mathcal{H}_a=(\bm{\alpha}\mathbf{p})+\beta m_a-\mu_a \beta 
(\bm{\Sigma}\mathbf{B})$ is the Hamiltonian for the particle $\psi_a$ accounting 
for the magnetic field, $V=-\mu \beta (\bm{\Sigma}\mathbf{B})$ describes the 
interaction of the transition magnetic moment with the external magnetic field, 
$\mu_a=\mu_{aa}$, and $\mu=\mu_{12}=\mu_{21}$ are elements of the matrix 
$({\mu}_{ab})$.

To find the general solution to Eq.~\eqref{Direqpsi} we follow the method used in 
Sec.~\ref{SO} and introduce the new wave functions $\tilde{\psi}_a = 
\mathcal{U}^\dag\psi_a$. All the calculations are identical to those made in 
Sec.~\ref{SO}. Therefore we present the final result for the wave functions 
$\tilde{\psi}_a$,
\begin{align}\label{GsolDPeq}
  \tilde{\psi}_{a}(x,t)= &
  \int_{-\infty}^{+\infty} \frac{\mathrm{d}p}{\sqrt{2\pi}}
  e^{\mathrm{i} p x}
  \notag
  \\
  & \times
  \sum_{\zeta=\pm 1}
  \Big[
    a_a^{(\zeta)}(t)u_a^{(\zeta)}\exp{(-\mathrm{i}E_a^{(\zeta)} t)}
	\notag
	\\
	& +
    b_a^{(\zeta)}(t)v_a^{(\zeta)}\exp{(+\mathrm{i}E_a^{(\zeta)} t)}
  \Big],
\end{align}
where the energy levels $E_a^{(\zeta)}$ are
\begin{equation}\label{energyDM}
  E_a^{(\zeta)}=
  \sqrt{\mathcal{M}_a^2 + m_a^2 + p^2 - 2 \zeta R_a^2}.
\end{equation}
Here [see Eq.~\eqref{EnergymattB}]
\begin{align}\label{RaMa}
  R_a^2= & \sqrt{p^2 \mathcal{M}_a^2 + (\mu_a B)^2 m_a^2},
  \notag
  \\
  \mathcal{M}_a= & \sqrt{(\mu_a B)^2 + \omega^2/4}.
\end{align}
The basis spinors $u_a^{(\zeta)}$ and $v_a^{(\zeta)}$ can be obtained from 
Eq.~\eqref{spinorsmattB} by the following replacement: $\mu \to \mu_a$,
$\mathcal{M} \to \mathcal{M}_a$ and $f^0 \to 0$. Our main goal is to determine the 
non-operator coefficients $a_a^{(\zeta)}$ and $b_a^{(\zeta)}$ so that to satisfy 
both the initial condition~\eqref{inicondpsi} and the evolution 
equation~\eqref{Direqpsi}. Generally the coefficients $a_a^{(\zeta)}(t)$ and 
$b_a^{(\zeta)}(t)$ are functions of time.

\subsection{Spin-flavor oscillations in case of diagonal magnetic 
moments\label{DMM}}

In this section we suppose that magnetic moments matrix in the mass eigenstates 
basis is close to diagonal, i.e. $\mu_a \gg \mu$. This case should be analyzed with 
help of the perturbation theory. We expand the wave functions $\tilde{\psi}_a$ in a 
series 
\begin{equation}\label{expan}
  \tilde{\psi}_{a}(x,t)=
  \tilde{\psi}^{(0)}(x,t)+
  \tilde{\psi}_{a}^{(1)}(x,t)+\dots,
\end{equation}
where $\tilde{\psi}^{(0)}(x,t)$ corresponds to the solution of Eq.~\eqref{GsolDPeq} 
when we neglect the potential $V$ there. The function $\tilde{\psi}_{a}^{(1)}(x,t)$ 
is linear in the transition magnetic moment $\mu$ etc. We omit terms of higher 
order in $\mu$ in Eq.~\eqref{expan}. They can be also accounted for but the 
corresponding calculations arrear to be cumbersome in the general case.      

Using orthonormality conditions of the basis spinors [see also Eq.~\eqref{oncso}],
\begin{equation*}
  u_a^{(\zeta)\dag}u_a^{(\zeta')}=v_a^{(\zeta)\dag}v_a^{(\zeta')}=
  \delta_{\zeta\zeta'},
  \quad
  u_a^{(\zeta)\dag}v_a^{(\zeta')}=0,
\end{equation*}
and the results of our previous work~\cite{DvoMaa07} (see also Sec.~\ref{SO}) we 
can receive from Eq.~\eqref{GsolDPeq} the expression for the zero order (in $\mu$) 
wave functions $\psi_a^{(0)}$, which correspond to the first term in 
Eq.~\eqref{expan},
\begin{equation}\label{solpsi0}
  \psi_{a}^{(0)}(x,t)=
  \mathcal{U}(x)
  \int_{-\infty}^{+\infty} \frac{\mathrm{d} p}{2\pi}
  e^{\mathrm{i} p x}
  S_a(p,t)\tilde{\psi}_{a}(p,0).
\end{equation}
where
\begin{equation}
  \tilde{\psi}_a(p,0)=
  \int_{-\infty}^{+\infty}\mathrm{d}x
  e^{-\mathrm{i}px}\mathcal{U}^\dag(x)\psi_a(x,0),
\end{equation}
is the Fourier transform of the initial condition for the spinor $\tilde{\psi}_a$. 
Here
\begin{align}\label{PJfB}
  S_a(p,t)= &
  \sum_{\zeta=\pm 1}
  \bigg[
    \left(
      u_a^{(\zeta)}\otimes u_a^{(\zeta)\dag}
    \right)
    \exp{(-\mathrm{i}E_a^{(\zeta)} t)}
	\notag
	\\
	& +
    \left(
      v_a^{(\zeta)}\otimes v_a^{(\zeta)\dag}
    \right)
    \exp{(+\mathrm{i}E_a^{(\zeta)} t)}
  \bigg],
\end{align}
is the analog of the Pauli-Joudan function in the twisting magnetic field [see also 
Eq.~\eqref{PJmattB}].

Using Eqs.~\eqref{solpsi0}-\eqref{PJfB} for the given initial condition we can find 
the wave functions at any subsequent moments of time. For example, if one initially 
has the left-handed neutrino $\nu_\beta^\mathrm{L}$, then field distribution of the 
right-handed component of the fermion $\nu_\alpha$ is
\begin{align}\label{nualphaR0}
  \nu^{(0)\mathrm{R}}_\alpha(x,t) = &
  \frac{1}{2}(1+\Sigma_1)[\cos\theta\psi_1(x,t)-\sin\theta\psi_2(x,t)]
  \notag
  \\
  = & \sin\theta\cos\theta e^{\mathrm{i}(k+\omega)x}
  \\
  & \times 
  \bigg[
    \frac{\mu_1 B}{2\mathcal{M}_1}
    \left(
      e^{-\mathrm{i}E^{+{}}_1 t}-e^{-\mathrm{i}E^{-{}}_1 t}
    \right)
	\notag
	\\
	\notag
	& -
    \frac{\mu_2 B}{2\mathcal{M}_2}
    \left(
      e^{-\mathrm{i}E^{+{}}_2 t}-e^{-\mathrm{i}E^{-{}}_2 t}
    \right)
  \left.
  \bigg]
  \right|_{p=k+\omega/2}\kappa_0.
\end{align}
To receive Eq.~\eqref{nualphaR0} we use the same technique as in Sec.~\ref{SO}. 
Therefore we may omit the details of calculations. On the basis of 
Eq.~\eqref{nualphaR0} one obtains the transition probability for the process 
$\nu^\mathrm{L}_\beta \to \nu^\mathrm{R}_\alpha$ in the form  
\begin{align}\label{PtrLR0}
  P^{(0)}_{\nu^\mathrm{L}_\beta \to \nu^\mathrm{R}_\alpha}(t) = &
  \frac{\sin^2 (2\theta)}{4}
  \bigg\{
    \bigg[
	  \frac{\mu_1 B}{\mathcal{M}_1}
	  \sin
	  \left(
	    \frac{E_1^{+{}}-E_1^{-{}}}{2}t
	  \right)
	  \\
	  & -
	  \frac{\mu_1 B}{\mathcal{M}_1}
	  \sin
	  \left(
	    \frac{E_2^{+{}}-E_2^{-{}}}{2}t
	  \right)
    \bigg]^2
	\notag
	\\
	& +
	4 \frac{\mu_1 \mu_2 B^2}{\mathcal{M}_1\mathcal{M}_2}
	\sin
	\left(
	  \frac{E_1^{+{}}-E_1^{-{}}}{2}t
	\right)
	\notag
	\\
	& \times
	\sin
	\left(
	  \frac{E_2^{+{}}-E_2^{-{}}}{2}t
	\right)
	\notag
	\\
	\notag
	& \times
	\sin^2
	\left(
	  \frac{E_1^{+{}}+E_1^{-{}}-E_2^{+{}}-E_2^{-{}}}{4}t
	\right)
  \bigg\}.
\end{align}
The energies $E^{(\zeta)}_a$ in Eqs.~\eqref{nualphaR0} and~\eqref{PtrLR0} are given 
in Eq.~\eqref{energyDM}. In Eq.~\eqref{PtrLR0} we also suppose that $p=k+\omega/2$.

The analysis of Eq.~\eqref{PtrLR0} is almost identical to that in Sec.~\ref{SO}. 
Therefore we present in the explicit form the final results for the wave function 
and the transition probability in the most important case when $p \gg 
\max(m_a,\mathcal{M}_a)$. This situation corresponds to spin-flavor oscillations of 
ultrarelativistic neutrinos. Now the wave function of $\nu_\alpha$ becomes
\begin{align}\label{nualphaR0fin}
  \nu^{(0)\mathrm{R}}_\alpha(x,t) = &
  \mathrm{i} \sin \theta \cos \theta
  \exp[\mathrm{i}(k+\omega)x-\mathrm{i} p t]
  \\
  & \times
  \bigg[
    \frac{\mu_1 B}{\mathcal{M}_1}\exp
    \left(
      -\mathrm{i}\frac{m_1^2}{2p}t
    \right)
    \sin\mathcal{M}_1 t
	\notag
	\\
	\notag
	& -
    \frac{\mu_2 B}{\mathcal{M}_2}
    \exp
    \left(
      -\mathrm{i}\frac{m_2^2}{2p}t
    \right)
    \sin\mathcal{M}_2 t
  \left.	
  \bigg]
  \right|_{p=k+\omega/2}\kappa_0,
\end{align}
and the transition probability in Eq.~\eqref{PtrLR0} is
\begin{align}\label{PtrLR0fin}
  P^{(0)}_{\nu^\mathrm{L}_\beta \to \nu^\mathrm{R}_\alpha}(t) = &
  \frac{\sin^2 (2\theta)}{4}
  \\
  & \times 
  \bigg\{
    \left(
      \frac{\mu_1 B}{\mathcal{M}_1}\sin\mathcal{M}_1 t-
      \frac{\mu_2 B}{\mathcal{M}_2}\sin\mathcal{M}_2 t
    \right)^2
	\notag
	\\
	\notag
	& +
    4\frac{\mu_1 \mu_2 B^2}{\mathcal{M}_1\mathcal{M}_2}
	\sin\mathcal{M}_1 t \sin\mathcal{M}_2 t \sin^2[\Phi(k)]
  \bigg\},
\end{align}
In Eq.~\eqref{PtrLR0fin} we use the notation for the oscillations phase,
\begin{equation}\label{Phi}
  \Phi(k)=\frac{\delta m^2}{4(k+\omega/2)}
\end{equation}
and $\delta m^2 = m_1^2 - m_2^2$ is the mass squared difference. In deriving 
Eqs.~\eqref{nualphaR0fin} and~\eqref{PtrLR0fin} we use the analog of the energy 
expansion in Eq.~\eqref{energy4case}. 

Note that the phase of oscillations in Eq.~\eqref{Phi} depends on the frequency of 
the twisting magnetic field. It should be noted that, if we put $\omega=0$ in 
Eqs.~\eqref{PtrLR0fin} and~\eqref{Phi}, the transition probability coincides with 
that from our work~\cite{DvoMaa07} where we studied neutrino spin-flavor 
oscillations in the constant transversal magnetic field.

If we studied the special case of massive neutrinos having equal magnetic moments, 
$\mu_1=\mu_2=\mu_0$, we would obtain the expected result from 
Eq.~\eqref{PtrLR0fin}. Namely, Eq.~\eqref{PtrLR0fin} can be rewritten as $P=P_F 
P_S$, where $P_F = \sin^2(2\theta)\sin^2[\Phi(k)t]$ is the usual transition 
probability of flavor oscillations and 
\begin{equation}\label{Ptrspin}
  P_S=\frac{(\mu_0 B)^2}{\Omega_S^2}
  \sin^2(\Omega_S t),
\end{equation}
is the probability of spin oscillations between different polarization states 
within each mass eigenstate. In Eq.~\eqref{Ptrspin} $\Omega_S = \sqrt{(\mu_0 
B)^2+(\omega/2)^2}$. That is, since the magnetic moment interactions are 
insensitive to flavor, the transitions between flavors are solely due to the mass 
mixing.

One can obtain the first order corrections (linear in $\mu$) to 
Eqs.~\eqref{nualphaR0fin} and~\eqref{PtrLR0fin}. These corrections correspond to 
the second term in Eq.~\eqref{expan}. The expressions for the corrections to the 
mass eigenstates wave functions are
\begin{align}\label{solpsi1}
  \psi_{a}^{(1)}(x,t)= &
  -\mathrm{i}\mathcal{U}(x)
  \int_{-\infty}^{+\infty} \frac{\mathrm{d}p}{2\pi} 
  e^{\mathrm{i} p x}
  \\
  & \times
  \sum_{\zeta=\pm 1}
  \Big[
    \left(
	  u_a^{(\zeta)}\otimes u_a^{(\zeta)\dag}
	\right)
	\exp{(-\mathrm{i}E_a^{(\zeta)} t)}
	\tilde{V}\mathcal{G}_a^\mathrm{(\zeta)}
	\notag
	\\
	\notag
	& +
	\left(
	  v_a^{(\zeta)}\otimes v_a^{(\zeta)\dag}
	\right)
	\exp{(+\mathrm{i}E_a^{(\zeta)} t)}
	\tilde{V}\mathcal{R}_a^\mathrm{(\zeta)}
  \Big]
  \tilde{\psi}_{b}(p,0),
\end{align}
where
\begin{align}\label{PJfint}
  \mathcal{G}_a^{(\zeta)}= &
  \int_0^t \mathrm{d}t' \exp{(+\mathrm{i}E_a^{(\zeta)}t')}S_b(p,t'),
  \notag
  \\
  \mathcal{R}_a^{(\zeta)}= &
  \int_0^t \mathrm{d}t' \exp{(-\mathrm{i}E_a^{(\zeta)}t')}S_b(p,t').
\end{align}
In Eqs.~\eqref{solpsi1} and~\eqref{PJfint} $a\neq b$. For the details of the 
derivation of Eqs.~\eqref{solpsi1} and~\eqref{PJfint} the reader is referred to 
Ref.~\cite{DvoMaa07} and Sec.~\ref{SO} of the present paper. Note that $\tilde{V} = 
- \mu B \beta \Sigma_3$ in Eq.~\eqref{solpsi1}.

The calculations of the the first order corrections based on Eqs.~\eqref{solpsi1} 
and~\eqref{PJfint} are rather cumbersome. Therefore we present here only the final 
results in the case when $p \gg \max(m_a,\mathcal{M}_a)$. One has the expression 
for the correction to the wave function,  
\begin{widetext}
\begin{align}\label{nualphaR1fin}
  \nu^{(1)\mathrm{R}}_\alpha(x,t) & =
  \mathrm{i} \mu B e^{\mathrm{i}(k+\omega)x} 
  \frac{1}{4\mathcal{M}_1\mathcal{M}_2}
  \displaybreak[1]
  \bigg[
    (\mu_1 \mu_2 B^2 + \mathcal{M}_1 \mathcal{M}_2 - \omega^2/4) \cos 2 \theta
	\left(
      \frac{\sin \delta t}{\delta}e^{-\mathrm{i} \sigma t}+
      \frac{\sin \Delta t}{\Delta}e^{-\mathrm{i} \Sigma t}
    \right)
	\\
    & -
    (\mu_1 \mu_2 B^2 - \mathcal{M}_1 \mathcal{M}_2 - \omega^2/4) \cos 2 \theta
	\displaybreak[1]
	\left(
      \frac{\sin d t}{d}e^{-\mathrm{i} s t}+
      \frac{\sin D t}{D}e^{-\mathrm{i} S t}
    \right)
	\notag
    \\
	\notag
	\displaybreak[1]
    & +
    (\mathcal{M}_1 - \mathcal{M}_2)\frac{\omega}{2}
    \left(
      \frac{\sin \delta t}{\delta}e^{-\mathrm{i} \sigma t}-
      \frac{\sin \Delta t}{\Delta}e^{-\mathrm{i} \Sigma t}
    \right)-
    (\mathcal{M}_1 + \mathcal{M}_2)\frac{\omega}{2}
	\left(
      \frac{\sin d t}{d}e^{-\mathrm{i} s t}-
      \frac{\sin D t}{D}e^{-\mathrm{i} S t}
    \right)
  \left.
  \bigg]
  \right|_{p=k+\omega/2}\kappa_0.
\end{align}
In Eq.~\eqref{nualphaR1fin} we use the notations,
\begin{align*}
  \sigma = & \frac{E_1^{+{}} + E_2^{+{}}}{2} \approx
  p + \Upsilon(k) - \bar{\mathcal{M}},
  \quad
  s = \frac{E_1^{+{}} + E_2^{-{}}}{2} \approx
  p + \Upsilon(k) - \delta \mathcal{M},
  \\
  \Sigma = & \frac{E_1^{-{}} + E_2^{-{}}}{2} \approx
  p + \Upsilon(k) + \bar{\mathcal{M}},
  \quad
  S = \frac{E_1^{-{}} + E_2^{+{}}}{2} \approx
  p + \Upsilon(k) + \delta \mathcal{M},
\end{align*}
and
\begin{align*}
  \delta = & \frac{E_1^{+{}} - E_2^{+{}}}{2} \approx
  \Phi(k) - \delta \mathcal{M},
  \quad
  d = \frac{E_1^{+{}} - E_2^{-{}}}{2} \approx
  \Phi(k) - \bar{\mathcal{M}},
  \\
  \Delta = & \frac{E_1^{-{}} - E_2^{-{}}}{2} \approx
  \Phi(k) + \delta \mathcal{M},
  \quad
  D = \frac{E_1^{-{}} - E_2^{+{}}}{2} \approx
  \Phi(k) + \bar{\mathcal{M}},
\end{align*}
where 
\begin{equation*}
  \Upsilon(k)=\frac{m_1^2+m_2^2}{4(k+\omega/2)},
  \quad
  \delta \mathcal{M} = \frac{\mathcal{M}_1 - \mathcal{M}_2}{2},
  \quad
  \bar{\mathcal{M}} = \frac{\mathcal{M}_1 + \mathcal{M}_2}{2}.
\end{equation*}
To obtain Eq.~\eqref{nualphaR1fin} we use the identity $\langle v_a^{(\zeta)} | 
\tilde{V} | \xi_0 \rangle = 0$, which means that no antineutrinos are produced.

On the basis of Eqs.~\eqref{nualphaR0fin} and~\eqref{nualphaR1fin} one calculates 
the correction to the transition probability which has the following form:
\begin{align}\label{PtrLR1fin}
  P^{(1)}_{\nu^\mathrm{L}_\beta \to \nu^\mathrm{R}_\alpha}(t) = &
  \frac{\mu B \sin 2\theta}{4\mathcal{M}_1\mathcal{M}_2}
  \bigg\{
    (\mu_1 \mu_2 B^2 + \mathcal{M}_1 \mathcal{M}_2 - \omega^2/4)
    \frac{\cos 2 \theta}{Z_1}
	\notag
    \\
	\displaybreak[1]
    & \times
    \bigg[
      \Phi(k)\sin[2\Phi(k)t]
      \left(
        \frac{\mu_1 B}{\mathcal{M}_1}\sin \mathcal{M}_1 t \cos \mathcal{M}_2 t -
        \frac{\mu_2 B}{\mathcal{M}_2}\cos \mathcal{M}_1 t \sin \mathcal{M}_2 t
      \right)
	  \notag
      \\
	  \displaybreak[1]
      & -
      \delta \mathcal{M}
      \bigg(
        \frac{\mu_1 B}{\mathcal{M}_1}\sin^2 \mathcal{M}_1 t +
        \frac{\mu_2 B}{\mathcal{M}_2}\sin^2 \mathcal{M}_2 t -
        \left(
          \frac{\mu_1 B}{\mathcal{M}_1}+\frac{\mu_2 B}{\mathcal{M}_2}
        \right)
        \sin \mathcal{M}_1 t \sin \mathcal{M}_2 t
		\notag
        \\
        & + 2
        \left(
          \frac{\mu_1 B}{\mathcal{M}_1}+\frac{\mu_2 B}{\mathcal{M}_2}
        \right)
        \sin \mathcal{M}_1 t \sin \mathcal{M}_2 t \sin^2[\Phi(k)t]
      \bigg)
    \bigg]
	\notag
    \\
	\displaybreak[1]
    & -
    (\mu_1 \mu_2 B^2 - \mathcal{M}_1 \mathcal{M}_2 - \omega^2/4)
    \frac{\cos 2 \theta}{Z_2}
	\bigg[
      \Phi(k)\sin[2\Phi(k)t]
      \left(
        \frac{\mu_1 B}{\mathcal{M}_1}\sin \mathcal{M}_1 t \cos \mathcal{M}_2 t -
        \frac{\mu_2 B}{\mathcal{M}_2}\cos \mathcal{M}_1 t \sin \mathcal{M}_2 t
      \right)
	  \notag
      \\
	  \displaybreak[1]
      & -
      \bar{\mathcal{M}}
      \bigg(
        \frac{\mu_1 B}{\mathcal{M}_1}\sin^2 \mathcal{M}_1 t -
        \frac{\mu_2 B}{\mathcal{M}_2}\sin^2 \mathcal{M}_2 t +
        \left(
          \frac{\mu_1 B}{\mathcal{M}_1}-\frac{\mu_2 B}{\mathcal{M}_2}
        \right)
        \sin \mathcal{M}_1 t \sin \mathcal{M}_2 t
		\notag
        \\
		\displaybreak[1]
        & - 2
        \left(
          \frac{\mu_1 B}{\mathcal{M}_1}-\frac{\mu_2 B}{\mathcal{M}_2}
        \right)
        \sin \mathcal{M}_1 t \sin \mathcal{M}_2 t \sin^2[\Phi(k)t]
      \bigg)
    \bigg]
	\notag
    \\
	\displaybreak[1]
    & +
    (\mathcal{M}_1 - \mathcal{M}_2)
    \frac{\omega}{2 Z_1}
    \bigg[
      \delta \mathcal{M} \sin[2\Phi(k)t]
      \left(
        \frac{\mu_1 B}{\mathcal{M}_1}\sin \mathcal{M}_1 t \cos \mathcal{M}_2 t -
        \frac{\mu_2 B}{\mathcal{M}_2}\cos \mathcal{M}_1 t \sin \mathcal{M}_2 t
      \right)
	  \notag
      \\
	  \displaybreak[1]
      & -
      \Phi(k)
      \bigg(
        \frac{\mu_1 B}{\mathcal{M}_1}\sin^2 \mathcal{M}_1 t +
        \frac{\mu_2 B}{\mathcal{M}_2}\sin^2 \mathcal{M}_2 t -
        \left(
          \frac{\mu_1 B}{\mathcal{M}_1}+\frac{\mu_2 B}{\mathcal{M}_2}
        \right)
        \sin \mathcal{M}_1 t \sin \mathcal{M}_2 t
		\notag
        \\
		\displaybreak[1]
        & + 2
        \left(
          \frac{\mu_1 B}{\mathcal{M}_1}+\frac{\mu_2 B}{\mathcal{M}_2}
        \right)
        \sin \mathcal{M}_1 t \sin \mathcal{M}_2 t \sin^2[\Phi(k)t]
      \bigg)
    \bigg]
	\notag
    \\
	\displaybreak[1]
    & -
    (\mathcal{M}_1 + \mathcal{M}_2)
    \frac{\omega}{2 Z_2}
    \bigg[
      \bar{\mathcal{M}}\sin[2\Phi(k)t]
      \left(
        \frac{\mu_1 B}{\mathcal{M}_1}\sin \mathcal{M}_1 t \cos \mathcal{M}_2 t -
        \frac{\mu_2 B}{\mathcal{M}_2}\cos \mathcal{M}_1 t \sin \mathcal{M}_2 t
      \right)
	  \notag
      \\
	  \displaybreak[1]
      & -
      \Phi(k)
      \bigg(
        \frac{\mu_1 B}{\mathcal{M}_1}\sin^2 \mathcal{M}_1 t -
        \frac{\mu_2 B}{\mathcal{M}_2}\sin^2 \mathcal{M}_2 t +
        \left(
          \frac{\mu_1 B}{\mathcal{M}_1}-\frac{\mu_2 B}{\mathcal{M}_2}
        \right)
        \sin \mathcal{M}_1 t \sin \mathcal{M}_2 t
		\notag
        \\
		\displaybreak[1]
        & - 2
        \left(
          \frac{\mu_1 B}{\mathcal{M}_1}-\frac{\mu_2 B}{\mathcal{M}_2}
        \right)
        \sin \mathcal{M}_1 t \sin \mathcal{M}_2 t \sin^2[\Phi(k)t]
      \bigg)
    \bigg]
  \bigg\},
\end{align}
\end{widetext}
where
\begin{equation}\label{Z1Z2}
  Z_1 = \Phi^2(k) - \delta \mathcal{M}^2,
  \quad
  Z_2 = \Phi^2(k) - \bar{\mathcal{M}}^2.
\end{equation}
Note that, if we again put $\omega=0$ in Eqs.~\eqref{nualphaR1fin} 
and~\eqref{PtrLR1fin}, we reproduce the results of our previous 
work~\cite{DvoMaa07}.

It can be noticed from Eqs.~\eqref{PtrLR1fin} and~\eqref{Z1Z2} that the 
perturbative approach is valid until $Z_{1,2} \neq 0$. If either $Z_1$ or $Z_2$ is 
equal to zero, we can expect that some non-perturbative effects like resonances can 
occur. Unfortunately these effects cannot be quantitatively described in frames of 
the approach based on the Dirac-Pauli equation used in the present work. To analyze 
such phenomena one should carry out numerical computations within the Schr\"odinger 
equation approach (see also Sec.~\ref{QM}). Nevertheless we evaluate the 
possibility that $Z_1=0$ for spin-flavor oscillations between active and sterile 
neutrinos (see, e.g., Ref.~\cite{KerMaaMyyRii04}) in the twisting magnetic field of 
the Sun. In this case one can take into account the magnetic moment of an active 
neutrino only. For the following parameters: $k \sim 10\thinspace\text{MeV}$, 
$\delta m^2 \sim 10^{-8}\thinspace\text{eV}^2$~\cite{KerMaaMyyRii04}, 
$\mu_\mathrm{activ} \sim 10^{-11} \mu_\mathrm{B}$, $B \sim 100\thinspace\text{kG}$ 
and $\omega \sim 10^{-15}\thinspace\text{eV}$~\cite{twisting}, we can see that the 
quantities $\mu_\mathrm{activ} B$, $\Phi(k)$ and $\omega$ are of the same order of 
magnitude of $10^{-15}\thinspace\text{eV}$. Thus the violation of the validity of 
the perturbation theory is quite possible for this kind of situation and resonance 
phenomena can happen.    

The sum of Eqs.~\eqref{PtrLR0fin} and~\eqref{PtrLR1fin} gives one the transition 
probability of spin-flavor oscillations up to terms linear in $\mu$ in case of  the 
magnetic moments matrix which is close to diagonal.

\subsection{Spin-flavor oscillations in case of non-diagonal magnetic 
moments\label{MMM}}

In this section we study neutrino spin-flavor oscillations in case of the 
non-diagonal magnetic moments matrix. It means that the transition magnetic moment 
dominates over the diagonal ones, i.e. we assume that $\mu \gg \mu_a$.

Now we start directly from Eq.~\eqref{GsolDPeq}. However one cannot treat the 
potential $\tilde{V}$ as the small perturbation. To solve this problem we should 
use the method elaborated in Ref.~\cite{DvoMaa07}. The following ordinary 
differential equations can be derived to determine the coefficients 
$a_a^{(\zeta)}(t)$ and $b_a^{(\zeta)}(t)$ in Eq.~\eqref{GsolDPeq}:
\begin{align}\label{abEqsGc}
  \mathrm{i}\dot{a}_a^{(\zeta)}= & 
  \exp{(+\mathrm{i}E_a^{(\zeta)}t)}u^{(\zeta)\dag}
  \tilde{V}
  \notag
  \\
  & \times
  \sum_{\zeta'=\pm 1}
  \Big[
    a_b^{(\zeta')}u^{(\zeta')}\exp{(-\mathrm{i}E_b^{(\zeta')} t)}
	\notag
    \\
    & +
	b_b^{(\zeta')}v^{(\zeta')}\exp{(+\mathrm{i}E_b^{(\zeta')} t)}
  \Big],
  \notag
  \\
  \mathrm{i}\dot{b}_a^{(\zeta)}= & 
  \exp{(-\mathrm{i}E_a^{(\zeta)} t)}v^{(\zeta)\dag}
  \tilde{V}
  \notag
  \\
  & \times
  \sum_{\zeta'=\pm 1}
  \Big[
    a_b^{(\zeta')}u^{(\zeta')}\exp{(-\mathrm{i}E_b^{(\zeta')} t)}
	\notag
    \\
    & +
	b_b^{(\zeta')}v^{(\zeta')}\exp{(+\mathrm{i}E_b^{(\zeta')} t)}
  \Big],
\end{align}
where $u^{+{}} = \kappa_0$, $u^{-{}} = \xi_0$, $v^{+\mathrm{T}} = (1/2)(1,1,-1,-1)$ 
and $v^{-\mathrm{T}} = (1/2)(1,-1,1,-1)$. The quantum number $\zeta$ is the 
eigenvalue of the operator $(\bm{\Sigma}\mathbf{p})/|\mathbf{p}|=\Sigma_1$, i.e. 
for example, $\Sigma_1 u^{(\zeta)} = \zeta u^{(\zeta)}$. Note that the current 
definition of $\zeta$ differs from that we use in previous sections. We also drop 
the subscripts $a$ and $b$ in the basis spinors since we study the evolution of 
ultrarelaticistic neutrinos and assume that $\mu_a \ll \mu$. The energies 
$E_a^{(\zeta)}$ in Eq.~\eqref{abEqsGc} take the form
\begin{equation}\label{energyMM}
  E_a^{(\zeta)}=\sqrt{m_a^2 + (p + \zeta \omega/2)^2}.
\end{equation}  
For the details of the derivation of Eq.~\eqref{energyMM} and the basis spinors the 
reader is referred to Sec.~\ref{SO} of this work.

The initial conditions should be added to the differential 
equations~\eqref{abEqsGc},
\begin{align}\label{abinicond}
  a_a^{(\zeta)}(0)=
  \frac{1}{\sqrt{2\pi}}u^{(\zeta)\dag}\tilde{\psi}_a(p,0),
  \notag
  \\
  b_a^{(\zeta)}(0)=
  \frac{1}{\sqrt{2\pi}}v^{(\zeta)\dag}\tilde{\psi}_a(p,0).
\end{align}
Eq.~\eqref{abinicond} results from Eq.~\eqref{GsolDPeq} and the orthonormality of 
the basis spinors $u^{(\zeta)}$ and $v^{(\zeta)}$.

Taking into account the following identities: $\langle u^{(\zeta)} | \tilde{V} | 
v^{(\zeta')} \rangle = 0$, $\langle u^{\pm{}} | \tilde{V} | u^{\pm{}} \rangle = 0$ 
and $\langle u^{\pm{}} | \tilde{V} | u^{\mp{}} \rangle = - \mu B$, which can be 
verified by means of direct calculations, one reveals that Eq.~\eqref{abEqsGc} is 
reduced to the form   
\begin{equation}\label{abEqsGcred}
  \mathrm{i}\dot{a}_a^{\pm{}} =
  -{a}_b^{\mp{}} \mu B \exp{[\mathrm{i}(E_a^{\pm{}}-E_b^{\mp{}})t]}.
\end{equation}
Note that the analogous equation for the functions $b_a^{(\zeta)}$ can be also 
obtained from Eq.~\eqref{abEqsGc}. Eq.~\eqref{abEqsGcred} is similar to that 
analyzed in Ref.~\cite{DvoMaa07} (see also Ref.~\cite{twisting}). Therefore we 
write down its solution, e.g., for the functions $a_a^{+{}}$,
\begin{equation}\label{a+MMsol}
  a_{1,2}^{+{}}(t)=
  \mathrm{i}\frac{\mu B}{\Omega_{\pm{}}} 
  \sin \Omega_{\pm{}} t 
  \exp(\pm \mathrm{i} \omega_{\pm{}} t/2) a_{1,2}^{-{}}(0),
\end{equation}
where 
\begin{equation}\label{Omegapmomegapm}
  \Omega_{\pm{}} = \sqrt{(\mu B)^2 + (\omega_{\pm{}}/2)^2},
  \quad
  \omega_{\pm{}} = 2\Phi(k) \pm \omega.
\end{equation}
In deriving Eq.~\eqref{a+MMsol} we take into account that initially only 
left-handed neutrinos are presented, i.e. $a_a^{+{}}(0)=0$. Indeed 
$\tilde{\psi}_a(p,0) \sim \xi_0$ and with help of Eq.~\eqref{abinicond} we get that 
$a_a^{+{}}(0)=0$. 

Finally, using Eqs.~\eqref{matrtrans}, \eqref{matrU}, \eqref{GsolDPeq} 
and~\eqref{a+MMsol} we arrive to the right-handed component of $\nu_\alpha$, 
\begin{align}\label{nualphaRMM}
  \nu_\alpha^\mathrm{R}(x,t)= &
  \mathrm{i} \mu B e^{\mathrm{i}(k+\omega)x}
  \\
  \notag
  \times &
  \exp
  \left[
    -\mathrm{i}
    \left(
      p + \frac{m_1^2+m_2^2}{4p}
    \right)t
  \right]
  \\
  \notag
  \times & 
  \left.
  \left(
    \cos^2\theta \frac{\sin \Omega_{+{}} t}{\Omega_{+{}}}-
    \sin^2\theta \frac{\sin \Omega_{-{}} t}{\Omega_{-{}}}  	
  \right)
  \right|_{p = k + \omega/2}
  \kappa_0.
\end{align}
With help of Eq.~\eqref{nualphaRMM} we can compute the transition probability for 
the process like $\nu^\mathrm{L}_\beta \to \nu^\mathrm{R}_\alpha$ in case of 
magnetic moments matrix with great non-diagonal elements,
\begin{align}\label{PtrLRMM}
  P_{\nu^\mathrm{L}_\beta \to \nu^\mathrm{R}_\alpha}(t)= &
  (\mu B)^2
  \\
  \notag
  & \times
  \left[
    \cos^2\theta \frac{\sin \Omega_{+{}} t}{\Omega_{+{}}}-
    \sin^2\theta \frac{\sin \Omega_{-{}} t}{\Omega_{-{}}}  	
  \right]^2.
\end{align}
Note that, if we approach to the limit $\omega=0$ in Eq.~\eqref{PtrLRMM}, we 
reproduce the result of our work~\cite{DvoMaa07}, where spin-flavor oscillations of 
neutrinos with similar magnetic moments matrix were studied.

Let us analyze neutrino oscillations at small frequencies of the twisting magnetic 
field, $\omega \ll \Omega_0 = \sqrt{(\mu B)^2 + [\Phi(k)]^2}$. In this situation 
the transition probability in Eq.~\eqref{PtrLRMM} can be rewritten as
\begin{align}\label{beating}
  P(t)= & A(t)\sin^2(\Omega_0 t),
  \notag
  \\
  A(t)= & A_\mathrm{min}+2\delta A\sin^2(\delta \Omega t).
\end{align}
where $\delta A = (A_\mathrm{max}-A_\mathrm{min})/2$, $\delta \Omega = 
(\Omega_{+{}}-\Omega_{-{}})/2 \approx \omega \Phi(k)/(2\Omega_0)$ and
\begin{align}\label{Amaxmin}
  A_\mathrm{max} & \approx 
  \left(
    \frac{\mu B}{\Omega_0}
  \right)^2
  \left[
    1-\frac{\omega \Phi(k)}{\Omega_0^2}\cos 2 \theta
  \right],
  \notag
  \\
  A_\mathrm{min} & \approx 
  \cos 2 \theta
  \left(
    \frac{\mu B}{\Omega_0}
  \right)^2
  \left[
    \cos 2 \theta-\frac{\omega \Phi(k)}{\Omega_0^2}
  \right].
\end{align}
Eqs.~\eqref{beating} and~\eqref{Amaxmin} show that the behavior of the system is 
analogous to beatings occurring in interference of two oscillations with different 
amplitudes and frequencies. 

\begin{figure}
  \centering
  \includegraphics[scale=.44]{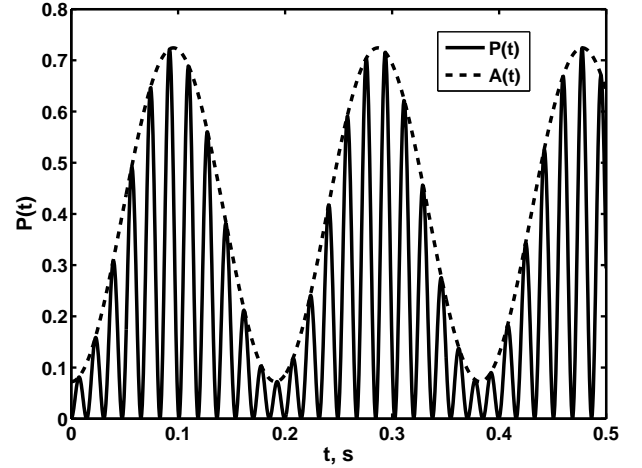}
  \caption{\label{figbeat}
  The time dependence of the neutrino oscillations 
  probability in the twisting magnetic
  field in the case $\mu\gg\mu_{1,2}$ at small values of $\omega$.}
\end{figure}

Let us discuss neutrinos with the following parameters: $\sin^2\theta = 0.3$, 
$\delta m^2 = 10^{-5}\thinspace\text{eV}^2$~\cite{MalSchTorVal04}, $\mu = 
10^{-18}\thinspace \mu_\mathrm{B}$ and $k=100\thinspace\text{MeV}$. It is known 
that rather strong twisting magnetic fields, up to the critical value of $B \sim 
10^{14}\thinspace\text{G}$, can exist in the early Universe~\cite{Ath96}. The time 
dependence of neutrino oscillations probability is schematically depicted on 
Fig.~\ref{figbeat} for such a magnetic field strength and $\omega = 
10^{-13}\thinspace\text{eV}$. 
As one can see on this figure, the rapidly varying transition probability $P(t)$ 
(solid line) is modulated by the slowly varying function $A(t)$ (dashed line). This 
time dependence is different from that described in Ref.~\cite{twisting}. 

It is also possible to see on Fig.~\ref{figbeat} that the typical time scale of the 
amplitude modulation of the transition probability is about $T_A \approx 
0.1\thinspace\text{s}$. The production rate of right-handed neutrinos in the early 
Universe should be less than the expansion rate of the Universe in order not to 
affect the primordial nucleosynthesis~\cite{prodrnuR}. Hence one has $T_A h > 1$, 
where 
\begin{equation*}
  h \approx 1.24 \times 10^{3}
  \left(
    \frac{T_\mathrm{pl}}{100\thinspace\text{MeV}}
  \right)^2
  \thinspace\text{s}^{-1},
\end{equation*}
is the Hubble parameter~\cite{Wei72} and $T_\mathrm{pl}$ is the primordial plasma 
temperature. Supposing that neutrinos are at thermal equilibrium at 
$100\thinspace\text{MeV}$, i.e. $k \sim T_\mathrm{pl}$, we get that $T_A h \sim 100 
\gg 1$. 
  
Despite the mentioned above discrepancy it is interesting to compare the result of 
this section [Eq.~\eqref{PtrLRMM}] with the analogous transition probability 
formula for Majorana neutrinos~\cite{twisting} at small mixing angle $\theta \to 
0$. Note that in this situation magnetic moments matrices in Eqs.~\eqref{Lagrnu} 
and~\eqref{Lagrpsi} [or Eq.~\eqref{magmomme}] coincide. In this limit we obtain 
from Eq.~\eqref{PtrLRMM} the following transition probability:
\begin{align}\label{PtrMajorana}
  P_{\nu^\mathrm{L}_\beta \to \nu^\mathrm{R}_\alpha}(t)= &
  \frac{(\mu B)^2}{\Omega_M^2}
  \sin^2(\Omega_M t),
\end{align}
where $\Omega_M = \sqrt{(\mu B)^2+[\Phi(k)+\omega/2]^2}$. It can be seen that 
Eq.~\eqref{PtrMajorana} coincides with the transition probabilities derived in 
Ref.~\cite{twisting} where spin-flavor oscillations of Majorana neutrinos in 
twisting magnetic fields were studied on the basis of the quantum mechanical 
approach.

It should be also noticed that the transition probability in Eq.~\eqref{PtrLRMM} 
vanishes at high frequencies, $\omega \gg \max(m_a, \mu B)$, due to the dependence 
of the oscillations phase on $\omega$ [see Eq.~\eqref{Phi}]. This phenomenon was 
also mentioned in our paper~\cite{Dvo07YadFiz} in which we examined spin-flavor 
oscillations of Majorana neutrinos in rapidly varying external fields.

\section{Quantum mechanical description of neutrino oscillations in a twisting 
magnetic field\label{QM}}

In this section we demonstrate that the analog of the main results, yielded in 
Secs.~\ref{DMM} and~\ref{MMM} within the Dirac-Pauli equation approach, can be also 
obtained with help of the standard quantum mechanical treatment of spin-flavor 
oscillations. The consistency between these two approaches is discussed.   

To study neutrinos evolution in frames of the Schr\"odinger equation approach it is 
convenient to make the coordinates transformation. We assume that 
$\mathbf{k}=(0,0,k)$ and $\mathbf{B}=B(\sin \omega t,\cos \omega t,0)$. Now the 
Schr\"odinger equation and the effective Hamiltonian for the neutrinos \emph{mass} 
eigenstates have the form,
\begin{equation}\label{Scheq}
  \mathrm{i}\dot{\Psi}=H\Psi,
  \quad
  H=
  \begin{pmatrix}
    H_\mathrm{mass} & H_B \\
	H_B^\dag & H_\mathrm{mass} \
  \end{pmatrix},
\end{equation}
where $H_\mathrm{mass}=\mathrm{diag}(\mathcal{E}_1,\mathcal{E}_2)$, 
$\mathcal{E}_{1,2}=\sqrt{m_{1,2}^2+k^2}$ and
\begin{equation}\label{HB}
  H_B=-(B_x+\mathrm{i}B_y)
  \begin{pmatrix}
    \mu_1 & \mu \\
	\mu & \mu_2 \
  \end{pmatrix}=
  - \mathrm{i} (\mu_{ab}) B e^{-\mathrm{i} \omega t},
\end{equation}
where $\mu_{1,2}$ and $\mu$ are the elements of the magnetic moments 
matrix~\eqref{magmomme}. The neutrinos wave function has the following form: 
$\Psi^\mathrm{T}=(\psi_1^\mathrm{L}, \psi_2^\mathrm{L}, \psi_1^\mathrm{R}, 
\psi_2^\mathrm{R})$, where $\psi_{1,2}^\mathrm{L,R}$ are one-component functions. 
It can be seen that Eqs.~\eqref{Scheq} and~\eqref{HB} is the generalization, for 
the case of Dirac neutrinos, of the corresponding expressions used in 
Ref.~\cite{twisting}. The initial condition for the wave function $\Psi$ follows 
from Eq.~\eqref{inicondpsi},
\begin{equation}\label{inicondqm}
  \Psi^\mathrm{T}(0)=(\sin\theta,\cos\theta,0,0).
\end{equation}  

Let us make the matrix transformation,
\begin{align}\label{rotbas}
  \Psi = & \mathfrak{V} \widetilde{\Psi},
  \notag
  \\
  \mathfrak{V}= & \mathrm{diag}
  (e^{-\mathrm{i} \omega t/2}, e^{-\mathrm{i} \omega t/2}, 
  e^{\mathrm{i} \omega t/2}, e^{\mathrm{i} \omega t/2})
\end{align}
The Hamiltonian $\tilde{H}$ governing the time evolution of the modified wave 
function $\widetilde{\Psi}$ is presented in the form
\begin{equation}\label{Scheqmod}
  \tilde{H}=
  \begin{pmatrix}
    H_\mathrm{mass}-\widehat{\mathds{1}}\omega/2 & -\mathrm{i}(\mu_{ab})B \\
	\mathrm{i}(\mu_{ab})B & H_\mathrm{mass}+\widehat{\mathds{1}}\omega/2 \
  \end{pmatrix},
\end{equation}
where $\widehat{\mathds{1}}$ is the $2 \times 2$ unit matrix. The initial condition 
for $\widetilde{\Psi}$ coincides with that for $\Psi$ [see Eq.~\eqref{inicondqm}] 
due to the special form of the matrix $\mathfrak{V}$ in Eq.~\eqref{rotbas}.

Then we look for the solutions to the Schr\"odinger equation 
$\mathrm{i}\mathrm{d}\widetilde{\Psi}/\mathrm{d}t=\tilde{H}\widetilde{\Psi}$, with 
the Hamiltonian given in Eq.~\eqref{Scheqmod}, in the form $\widetilde{\Psi} \sim 
e^{- \mathrm{i} \lambda t}$. The secular equation for $\lambda$ is the forth order 
algebraic equation in general case. However it can be solved in two situations.

\subsection{Diagonal magnetic moments matrix}

In the case when $\mu = 0$ and $\mu_{1,2} \neq 0$ the roots of the secular equation 
are
\begin{equation}\label{DMMroots}
  \lambda_{a}^{\pm{}}= \mathcal{E}_a \pm \mathcal{M}_a,
  \quad
  a=1,2,
\end{equation} 
where $\mathcal{M}_a$ is defined in Eq.~\eqref{RaMa}. The basis spinors 
$u_a^{\pm{}}$, which are the eigenvectors of the Hamiltonian $\tilde{H}$ are 
expresed in the following way:
\begin{align}\label{DMMspinors}
  u_1^{+{}}= &
  \frac{1}{\sqrt{2\mathcal{M}_1}}
  \begin{pmatrix}
    -\mathrm{i}\mu_1 B/\mathcal{Z}_1 \\
	0 \\
	\mathcal{Z}_1 \\
    0 \
  \end{pmatrix},
  \notag
  \\
  u_1^{-{}}= &
  \frac{1}{\sqrt{2\mathcal{M}_1}}
  \begin{pmatrix}
    \mathcal{Z}_1 \\
	0 \\
	-\mathrm{i}\mu_1 B/\mathcal{Z}_1 \\
    0 \
  \end{pmatrix},
  \notag
  \\
  u_2^{+{}}= &
  \frac{1}{\sqrt{2\mathcal{M}_2}}
  \begin{pmatrix}
    0 \\
	-\mathrm{i}\mu_2 B/\mathcal{Z}_2 \\
	0 \\
    \mathcal{Z}_2 \
  \end{pmatrix},
  \notag
  \\
  u_2^{-{}}= &
  \frac{1}{\sqrt{2\mathcal{M}_2}}
  \begin{pmatrix}
    0 \\
	\mathcal{Z}_2 \\
	0 \\
    -\mathrm{i}\mu_2 B/\mathcal{Z}_2 \
  \end{pmatrix},
\end{align}
where $\mathcal{Z}_{1,2}=\sqrt{\mathcal{M}_{1,2}+\omega/2}$. Note that the vectors 
$u_a^{\pm{}}$ correspond to the eigenvalues $\lambda_a^{\pm{}}$.

The general solution to the Schr\"odinger evolution equation has the form,
\begin{equation}\label{DMMgensol}
  \widetilde{\Psi}(t)=\sum_{a=1,2} \sum_{\zeta=\pm 1}
  \alpha_a^{(\zeta)}u_a^{(\zeta)}\exp(-\mathrm{i}\lambda_a^{(\zeta)}t),
\end{equation}
where the coefficients $\alpha_a^{(\zeta)}$ should be chosen so that to satisfy the 
initial condition in Eq.~\eqref{inicondqm}. We choose these quantities as
\begin{align}\label{DMMcoeff}
  \alpha_1^{+{}} = & \sin\theta \frac{1}{\sqrt{2M_1}}
  \frac{\mathrm{i}\mu_1 B}{\mathcal{Z}_1},
  \quad
  \alpha_1^{-{}} = \sin\theta \frac{\mathcal{Z}_1}{\sqrt{2M_1}},
  \notag
  \\
  \alpha_2^{+{}} = & \cos\theta \frac{1}{\sqrt{2M_2}}
  \frac{\mathrm{i}\mu_2 B}{\mathcal{Z}_2},
  \quad
  \alpha_2^{-{}} = \sin\theta \frac{\mathcal{Z}_1}{\sqrt{2M_1}},
\end{align}
Then using Eqs.~\eqref{rotbas} and~\eqref{DMMroots}-\eqref{DMMcoeff} we get the 
right-polarized components of the wave function $\Psi$ in the form
\begin{align}\label{DMMpsiR}
  \psi_1^\mathrm{R}(t)= & \exp[-\mathrm{i}(\mathcal{E}_1-\omega/2)t]
  \sin\theta \sin(\mathcal{M}_1 t)  
  \frac{\mu_1 B}{\mathcal{M}_1},
  \notag
  \\
  \psi_2^\mathrm{R}(t)= & \exp[-\mathrm{i}(\mathcal{E}_2-\omega/2)t]
  \cos\theta \sin(\mathcal{M}_2 t)  
  \frac{\mu_2 B}{\mathcal{M}_2}.
\end{align}
Finally taking into account Eqs.~\eqref{matrtrans}, \eqref{matrU} 
and~\eqref{DMMpsiR} we arrive to the right-handed component of $\nu_\alpha$,
\begin{align}\label{DMMnualphaR}
  \nu_\alpha^\mathrm{R}(t)= & 
  \cos\theta \psi_1^\mathrm{R}(t)-\sin\theta \psi_2^\mathrm{R}(t)
  \notag
  \\
  = &  
  \sin\theta \cos\theta e^{\mathrm{i}\omega t/2}
  \bigg[
    \exp(-\mathrm{i}\mathcal{E}_1 t)\sin(\mathcal{M}_1 t)
	\frac{\mu_1 B}{\mathcal{M}_1}
	\notag
    \\
    & -
	\exp(-\mathrm{i}\mathcal{E}_2 t)\sin(\mathcal{M}_2 t)
	\frac{\mu_2 B}{\mathcal{M}_2}
  \bigg].
\end{align}
One can see from Eq.~\eqref{DMMnualphaR} that the expression for 
$\nu_\alpha^\mathrm{R}$ obtained in frames of the Schr\"odinger approach coincides 
(to within some irrelevant phase factor) with the analogous expression derived 
using the Dirac-Pauli equation [see Eq.~\eqref{nualphaR0fin}]. 

\subsection{Non-diagonal magnetic moments matrix}

In the situation when $\mu_{1,2}=0$ and $\mu \neq 0$ the secular equation can be 
also solved analytically and the corresponding roots are
\begin{equation}\label{MMMroots}
  \lambda_{1,2}^{+{}}= \bar{\mathcal{E}} \pm \Omega_{+{}},
  \quad
  \lambda_{1,2}^{-{}}= \bar{\mathcal{E}} \pm \Omega_{-{}},
\end{equation} 
where $\bar{\mathcal{E}}=(\mathcal{E}_1+\mathcal{E}_2)/2$ and $\Omega_{\pm{}}$ are 
given in Eq.~\eqref{Omegapmomegapm}. The eigenvectors of the Hamiltonian 
$\tilde{H}$ have the following form:
\begin{align}\label{MMMspinors}
  u_{1,2} = &
  \frac{1}{\sqrt{2\Omega_{+{}}}}
  \begin{pmatrix}
    0 \\
	\mp \mathrm{i}\mu B/R_{\pm{}} \\
	R_{\pm{}} \\
    0 \
  \end{pmatrix},
  \notag
  \\
  v_{1,2} = &
  \frac{1}{\sqrt{2\Omega_{-{}}}}
  \begin{pmatrix}
    \mp \mathrm{i}\mu B/S_{\mp{}} \\
	0 \\
	0 \\
    S_{\mp{}} \
  \end{pmatrix},
\end{align}     
where $R_{\pm{}}=\sqrt{\Omega_{+{}} \pm \omega_{+{}}/2}$, 
$S_{\pm{}}=\sqrt{\Omega_{-{}} \pm \omega_{-{}}/2}$ and $\omega_{\pm{}}$ are given 
in Eq.~\eqref{Omegapmomegapm}. The spinors $u_{1,2}$ and $v_{1,2}$ in 
Eq.~\eqref{MMMspinors} correspond to the eigenvalues $\lambda_{1,2}^{+{}}$ and 
$\lambda_{1,2}^{-{}}$ respectively.

The general solution to the Schr\"odinger equation for the function 
$\widetilde{\Psi}$ takes the form,
\begin{align}\label{MMMgensol}
  \widetilde{\Psi}(t)= & \exp(-\mathrm{i}\bar{\mathcal{E}}t)
  [\alpha_1 u_1 \exp(-\mathrm{i}\Omega_{+{}}t)+ 
  \alpha_2 u_2 \exp(\mathrm{i}\Omega_{+{}}t)
  \notag
  \\
  & +
  \beta_1 v_1 \exp(-\mathrm{i}\Omega_{-{}}t)+ 
  \beta_2 v_2 \exp(\mathrm{i}\Omega_{-{}}t)],
\end{align}
We again choose the coefficients $\alpha_{1,2}$ and $\beta_{1,2}$ in 
Eq.~\eqref{MMMgensol} to satisfy the initial condition in Eq.~\eqref{inicondqm}. 
These coefficients have to be chosen as
\begin{equation}\label{MMMcoeff}
  \alpha_{1,2} = \pm \mathrm{i} \cos\theta  
  \frac{R_{\mp{}}}{\sqrt{2\Omega_{+{}}}},
  \quad
  \beta_{1,2} = \pm \mathrm{i} \sin\theta 
  \frac{S_{\pm{}}}{\sqrt{2\Omega_{-{}}}},
\end{equation}
With help of Eqs.~\eqref{rotbas} and~\eqref{MMMroots}-\eqref{MMMcoeff} we obtain 
the right-handed components of the wave function $\Psi$ in the form
\begin{align}\label{MMMpsiR}
  \psi_1^\mathrm{R}(t)= & \mu B \cos\theta 
  \exp[-\mathrm{i}(\bar{\mathcal{E}}-\omega/2)t]
  \frac{\sin(\Omega_{+{}}t)}{\Omega_{+{}}},
  \notag
  \\
  \psi_2^\mathrm{R}(t)= & \mu B \sin\theta 
  \exp[-\mathrm{i}(\bar{\mathcal{E}}-\omega/2)t]
  \frac{\sin(\Omega_{-{}}t)}{\Omega_{-{}}}.
\end{align}
On the basis of Eq.~\eqref{MMMpsiR} we receive the wave function 
$\nu_\alpha^\mathrm{R}$ as
\begin{align}\label{MMMnualphaR}
  \nu_\alpha^\mathrm{R}(t) = &
  \mu B \exp[-\mathrm{i}(\bar{\mathcal{E}}-\omega/2)t]
  \notag
  \\
  & \times
  \left(
    \cos^2\theta \frac{\sin \Omega_{+{}} t}{\Omega_{+{}}}-
    \sin^2\theta \frac{\sin \Omega_{-{}} t}{\Omega_{-{}}}  	
  \right).
\end{align}
Comparing Eq.~\eqref{MMMnualphaR} with the analogous expression~\eqref{nualphaRMM} 
derived in frames of the Dirac-Pauli equation approach we again find an agreement 
to within the phase factor. 

It should be however mentioned that using the quantum mechanical treatment of 
spin-flavor oscillations one cannot reproduce the expression for the phase of the 
neutrino oscillations~\eqref{Phi}. It means that in Eqs.~\eqref{DMMnualphaR} 
and~\eqref{MMMnualphaR} we have the standard quantum mechanical vacuum oscillations 
phase $\Phi_{QM}(k)=\delta m^2/(4k)$. 

\section{Applications\label{APPL}}

Let us discuss the applicability of our results to one specific oscillation  
channel, $\nu_\mu^\mathrm{L}\xleftrightarrow{B}\nu_\tau^\mathrm{R}$.
According to the recent neutrino oscillations data (see, e.g., 
Ref.~\cite{MalSchTorVal04}) the mixing angle between $\nu_\mu$ and $\nu_\tau$ is 
close to its maximal value of  $\pi/4$. For such a mixing angle the magnetic moment 
matrix given in Eq.~\eqref{magmomme} is expressed in the form,
\begin{multline}
  ({\mu}_{ab})\approx
  \\
  \begin{pmatrix}
    [{M}_{\tau\tau}+{M}_{\mu\mu}]/2+{M}_{\tau\mu} &
    -[{M}_{\tau\tau}-{M}_{\mu\mu}]/2 \\
    -[{M}_{\tau\tau}-{M}_{\mu\mu}]/2 &
    [{M}_{\tau\tau}+{M}_{\mu\mu}]/2-{M}_{\tau\mu} \
  \end{pmatrix}.
\end{multline}
Eq.~\eqref{PtrLR0fin} is valid when this matrix is close to diagonal, i.e. if 
$|({M}_{\tau\tau}-{M}_{\mu\mu})/2| \ll |({M}_{\tau\tau}+{M}_{\mu\mu})/2 
\pm{M}_{\tau\mu}|$.  In contrast to the mixing angles, experimental data and 
theoretical predictions for the values of neutrino magnetic moments are not very 
reliable~\cite{magnmomDM}. However it is known that the diagonal magnetic moments 
$M_{\lambda\lambda}$ could be very small in the extensions of the standard model 
${M}_{\lambda\lambda}\sim 
10^{-19}({m}_{\lambda\lambda}/\text{eV})\mu_\mathrm{B}$~\cite{nuMM}. The transition 
magnetic moments, ${M}_{\tau\mu}$ in our notations, can be much greater up to the 
experimental limit of $10^{-10}\mu_\mathrm{B}$~\cite{Yao06}. One can see that for 
any conceivable values of the masses of the known neutrinos,
${M}_{\mu\mu}$ and ${M}_{\tau\tau}$ are several orders of magnitude smaller than 
$10^{-10}\mu_\mathrm{B}$. Our result~\eqref{PtrLR0fin} is valid in this case.

It is worth mentioning that Eqs.~\eqref{solpsi0} and~\eqref{solpsi1} are in 
principle applicable for particles with arbitrary initial conditions, e.g., with 
small initial momenta. Therefore one can discuss the evolution and oscillations of 
relic neutrinos. These particles can gravitationally cluster in the 
Galaxy~\cite{RinWon05}. It is possible to relate the temperatures of relic 
neutrinos and CMB protons, $T_\nu$ and $T_\gamma$ respectively,
\begin{equation}
  T_\nu = 
  \left(
    \frac{4}{11}
  \right)^{1/3}
  T_\gamma
  \approx 0.72 T_\gamma.
\end{equation}
For the present value $T_\gamma = 2.7\ensuremath{^\circ}\thinspace\mathrm{K}$ we 
get $T_\nu \approx 1.93\ensuremath{^\circ}\thinspace\mathrm{K}$~\cite{RinWon05}. 
Using the estimate for the neutrino mass $m \sim 0.1\thinspace\mathrm{eV}$, because 
the sum of all neutrinos masses should be less than 
$1\thinspace\mathrm{eV}$~\cite{HanRaf06}, and taking into account that these 
neutrinos are non-relativistic particles, one obtains the typical momentum of a 
relic neutrino $k \sim 7 \times 10^{-3}\thinspace\mathrm{eV}$. For example, to 
realize the "non-standard" neutrino propagation regime described in 
item~\ref{case1} of Sec.~\ref{SO} one should use an undulator with the frequency 
$|\omega| \sim 2k$ or with the period $L = 2\pi/\omega \sim 
0.1\thinspace\mathrm{mm}$. 

Strong periodic electromagnetic fields, with a short spatial oscillations length, 
can be found in crystals~\cite{Ugg05}. In Ref.~\cite{Bel03} it was proposed to 
manufacture deformed crystals with submillimeter periods. The undulator radiation 
was recently reported in Ref.~\cite{Bar05} to be produced in undulators with 
periods of $0.1-1\thinspace\mathrm{mm}$. Therefore artificial  crystalline 
undulators with required periods are used in various experiments and hence they can 
serve as a good tool to explore properties of relic neutrinos. It should be 
mentioned that neutrino scattering on a polarized target and possible tests of 
neutrino magnetic moment were examined in Ref.~\cite{RasSem00}.                   

\section{Summary\label{CONCL}}

We have described the evolution of Dirac neutrinos in matter and in a twisting 
magnetic field. We have applied the recently developed approach (see 
Refs.~\cite{FOvac,Dvo06EPJC,DvoMaa07}) which is based on the exact solutions to the 
Dirac equation in an external field with the given initial condition.

First (Sec.~\ref{SO}) we have found the solution to the Dirac equation for a 
neutral $1/2$-spin particle weakly interacting with background matter, that is 
equivalent to an external axial-vector field, and non-minimally coupled to an 
external electromagnetic field due to the possible presence of an anomalous 
magnetic moment. We have discussed the situation when a neutrino interacts with the 
twisting magnetic field. The energy spectrum and basis spinors have been obtained. 
We have applied these results to derive the transition probability of spin 
oscillations in matter under the influence of the twisting magnetic field. The 
scope of the standard quantum mechanical approach to the description of neutrino 
spin oscillations has been analyzed. 

Then (Sec.~\ref{SFO}) we have used the obtained solution to the Dirac equation for 
the description of neutrino spin-flavor oscillations in a twisting magnetic field. 
We supposed that two Dirac neutrinos could mix and have non-vanishing matrix of 
magnetic moments. Moreover the mass and magnetic moments matrices in the flavor 
eigenstates basis are generally independent, i.e. the diagonalization of the mass 
matrix, that means the transition to the mass eigenstates basis, does not lead to 
the diagonalization of the magnetic moments matrix. We have discussed two 
possibilities. 

In Sec.~\ref{DMM} we have assumed that magnetic moments matrix in the mass 
eigenstates basis has great diagonal elements compared to the non-diagonal ones. In 
this case one can analyze neutrino spin-flavor oscillations perturbatively. Note 
that the perturbative approach allows one to discuss neutrinos with an arbitrary 
initial condition. For instance, the evolution of particles with small initial 
momenta can be accounted for. The appearance of non-perturbative phenomena like 
resonances is analyzed with an example of active-to-sterile neutrinos oscillations. 
We have discussed the opposite situation in Sec.~\ref{MMM}, i.e. the magnetic 
moment matrix with the great non-diagonal elements. In this case one had to treat 
the evolution of the system non-perturbatively. We have demonstrated that this 
situation is analogous to beatings resulting from the superposition of two 
oscillations. In both cases we have obtained neutrino wave functions, consistent 
with the initial conditions, and the transition probabilities. Note that all the 
results are in agreement with our previous work~\cite{DvoMaa07} if we set 
$\omega=0$, i.e. discuss a constant transversal magnetic field. We have also 
examined some limiting cases and compared our results with the previous studies.

It has been shown in Sec.~\ref{QM} that one can derive the analog of the major 
results obtained in Secs.~\ref{DMM} and~\ref{MMM} using the Schr\"odinger evolution 
equation approach for the description of spin-flavor oscillations of Dirac 
neutrinos. The correspondence between these two approaches has been considered.  

In Sec.~\ref{APPL} we have discussed magnetic moments matrices in various 
theoretical models which predict neutrinos magnetic moments. The validity of our 
approach for these situations has been considered. The applications of our results 
to the studying of cosmological neutrinos in laboratory conditions have been 
examined. In particular we have suggested that artificial crystalline undulators 
could be useful for such a research.     

The results obtained in the present work are valid for arbitrary magnetic field 
strength. The general case of spin-flavor oscillations of Dirac neutrinos in a 
twisting magnetic field with an arbitrary magnetic moment matrix has not been 
studied analytically earlier. Both experimental and theoretical information about 
magnetic moments of Dirac neutrinos is known to be very limited (see, e.g. 
Refs.~\cite{magnmomDM,Yao06}). Therefore our results can be helpful since they 
enable one to describe phenomenologically spin-flavor oscillations of Dirac 
neutrinos under the influence of the magnetic field in question provided neutrinos 
possess non-vanishing matrix of magnetic moments. Although we  consider neutrinos, 
it is possible to straightforwardly apply our formalism to the description of any 
$1/2$-spin particles.

\begin{acknowledgments}
The work has been supported by the Academy of Finland under the
contract No.~108875. The author is thankful to the Russian Science Support 
Foundation for a grant as well as to Efrain Ferrer (Western Illinois University) 
and Kimmo Kainulainen (University of Jyv\"askyl\"a) for useful comments. The 
referees' remarks are also appreciated.
\end{acknowledgments}

\end{document}